# Artificial Bee Colony optimization of Deep Convolutional Neural Networks in the context of Biomedical Imaging

A. Gómez[1,2], C. Fernández[1,2], M. Abella[1,2,3] and M. Desco[1,2,3,4]

[1] *Bioengineering Department, Universidad Carlos III, Madrid, España*
[2] *Instituto de Investigación Sanitaria Gregorio Marañón, Madrid, España*
[3] *Centro Nacional Investigaciones Cardiovasculares Carlos III (CNIC), Madrid, España*
[4] *Centro de Investigación Biomédica en Red de Salud Mental (CIBERSAM), Madrid, España*

*Abstract*—Most efforts in Computer Vision focus on natural images or artwork, which differ significantly —both in size and contents— from the kind of data biomedical image processing deals with. Thus, Transfer Learning models often prove themselves suboptimal for these tasks, even after manual fine-tuning. The development of architectures from scratch is oftentimes unfeasible due to the vastness of the hyperparameter space and a shortage of time, computational resources and Deep Learning experts in most biomedical research laboratories. An alternative to manually defining the models is the use of Neuroevolution, which employs metaheuristic techniques to optimize Deep Learning architectures. However, many algorithms proposed in the neuroevolutive literature are either too unreliable or limited to a small, predefined region of the hyperparameter space. To overcome these shortcomings, we propose the Chimera Algorithm, a novel, hybrid neuroevolutive algorithm that integrates the Artificial Bee Colony Algorithm with Evolutionary Computation tools to generate models from scratch, as well as to refine a given previous architecture to better fit the task at hand. The Chimera Algorithm has been validated with two datasets of natural and medical images, producing models that surpassed the performance of those coming from Transfer Learning.

## 1. Introduction

The development of Deep Learning tools is revolutionizing the healthcare sector. Their potential lays in their ability to automatically extract features and patterns from the training datasets and extrapolate them to process new, unseen data. Their versatility is reflected on their increasingly wide range of applications: from the analysis of omics data, the discovery and development of new drugs, the diagnosis of patients, or the processing and analysis of medical images [1]. However, the models developed for traditional Computer Vision problems are rarely optimal in medical imaging. Pretrained models, even when carefully fine-tuned, tend to underperform, likely because their hyperparameters and weights are tailored for tasks where the overall features hold much greater relevance than fine detail [2]. Moreover, the medical imaging data size tends to be much larger than in most natural images or artworks. Lastly, most Transfer Learning models pretrained on natural image datasets present a much higher number of outputs —for instance, ImageNet presents 1000 classes while most medical classification problems may need orders of magnitude less—, which leads to an unnecessary overparametrization of the latest layers.

One straightforward approach is to optimize the model's hyperparameters manually. However, there are no universal guidelines for the optimization of Convolutional Neural Networks, which makes it a complex and highly subjective endeavor that requires considerable time, resources and the expertise of trained data scientists. Moreover, the dimensionality of the set of hyperparameters grows exponentially with the model size: the larger the number of layers, the greater the number of combinations between their hyperparameters. This model size is not fixed, but rather another important hyperparameter to optimize.

One more difficulty is that there is no general method to predict the final model performance without training until convergence, which makes evaluating each candidate set of hyperparameters computationally expensive. Some efforts to approximate the final loss value early during training by means of fitting the learning curve [3] or exploring the connection dynamics of the system [4] have not reached ample acceptance.

Such a complex combinatorial problem can instead be automatically addressed by employing metaheuristic optimization, which gave rise to the field of Neuroevolution in the late 80s [5]. Genetic Algorithms and Evolutionary Computation are the most common metaheuristics to optimize either network weights, topology, or both simultaneously. These neuroevolutionary methods are attractive because they are problem-independent and can quickly explore the hyperparameter space [6, 7, 8]. However, they are frequently unable to reliably converge towards the best models locally and prone to converge prematurely due to a loss of diversity within the population as the search progresses [9]. This problem becomes more severe due to the unprecedented complexity of modern architectures. Newer methods, such as supernet optimization approaches [10] or Multi-Fidelity MetaLearning [11], intend to increase efficiency by limiting the search space, thus sacrificing exploration for the sake of exploitation.

The need for a compromise between exploration and exploitation led to the use of Swarm Intelligence metaheuristics, which offer the ability to share information throughout a population of optimizer agents in a cooperative fashion without a high computational cost [9]. Therefore, these optimizers are able to find and exploit regions of interest within the space more efficiently than competitive ones. In particular, Particle Swarm Optimization has already been successfully applied to the optimization of Convolutional Neural Networks [12, 13], outperforming Genetic and Evolutionary-based approaches.

However, Particle Swarm Optimization can only evolve architectures towards the best ones found so far. Therefore, it may be difficult to exploit neighboring solutions or reach architectures that are very different from the initialized population. These problems are overcome by Swarm Intelligence algorithms based on the coordinated exploitation of local regions in the solution space, such as the Artificial Bee Colony Algorithm [14]. This algorithm deploys a population of specialized agents in the solution space that perform random walk steps, focusing on the most promising regions. Its ability to freely exploit the neighborhood of any given solution allows starting the search from a single location provided by the user, using a-priori information to significantly speed up the search. In this work, we propose to adjust the Artificial Bee Colony Algorithm —which, to the best of our knowledge, has only been applied to numerical problems— to be used as a backbone for the neuroevolutive search of feedforward Convolutional Neural Networks. The neighboring solutions are explored by applying mutation operators developed for Evolutionary architecture search approaches. The resulting optimizer, referred to as the Chimera Algorithm, has been evaluated with a classification task on natural images and a regression task on Computed Tomography studies.

2. PROPOSED ALGORITHM

The Chimera Algorithm follows the same workflow as the Artificial Bee Colony Algorithm [14], as shown in Algorithm 1. It first initializes a population of solutions by copying a provided base architecture or by creating models with a random amount of layers, up to a maximum defined by the user or imposed by hardware limitation. This randomization can be done by either adding layers from scratch or mutating a specified architecture. Then, two types of optimizer agents, referred to as Employed and Onlooker Bees, will take turns to explore the hyperparameter space until certain specified stop criteria are met. First, each Employed Bee is bound to one solution in a one-to-one assignment and explore around by duplicating the solution and performing a series of mutation steps on the copy. We keep an exhaustion counter for each solution that increases every time the mutated model is worse than the original one, and resets to zero when it is not. If a solution were to become exhausted —that is, its exhaustion counter exceeds a certain threshold—, it is saved as a plausible global minimum and the Employed Bee reassigns itself to a newly initialized solution. The number of mutations performed on each iteration was defined as $|N(1, \sqrt[3]{1+c_e})|$ rounded upwards, where $c_e$ is the exhaustion counter for that solution, to ensure that most of the time we perform very few mutation steps. Bigger steps to overcome local minima are only allowed if the solution is close to exhaustion. In this way, we make sure that the close neighborhood of a given solution is properly exploited before trying to reach further away.

*Algorithm 1. Chimera Algorithm*

1. **Input:** *train_dset, val_dset, l_thresh, Np, max_iter, max_ex*
2. **Output:** population of *final_models*
3. Initialize a population of *models* of size *Np*
4. Each *model* is trained and validated
5. Each *model* is assigned a *score*
6. Initialize a population of *Employed_Bees* of size *Np*
7. Assign a *model* to each *Employed_Bee* ∈ *Employed_Bees*
8. Initialize a population of *Onlooker_Bees* of size *Np*
9. *final_models* ← ∅
10. **for** *i* = 0 to *max_iter* **do**
11.    **for** *E_Bee* ∈ *Employed_Bees* **do**
12.       *E_Bee* creates a *new_model* by adding, removing or modifying layers from its associated *model*
13.       The *new_model* is trained and assigned a *score*
14.       **if** *new_model*'s *score* < *E_Bee*'s *model*'s *score* **then**
15.          *E_Bee*'s *model* ← *new_model*
16.       **else**
17.          *E_Bee*'s *model*'s *exhaustion* += 1
18.       **if** *E_Bee*'s *model*'s *exhaustion* ≥ *max_ex* **then**
19.          *final_models* ← *final_models* ∪ *E_Bee*'s *model*
20.          *E_Bee*'s *model* is reinitialized
21.          *E_Bee*'s *model*'s *exhaustion* ← 0
22.    **for** *O_Bee* ∈ *Onlooker_Bees* **do**
23.       *O_Bee* randomly selects a *model*, weighted by score
24.       *O_Bee* creates a *new_model* by adding, removing or modifying layers from its selected *model*
25.       The *new_model* is trained and assigned a *score*
26.       **if** *new_model*'s score < *O_Bee*'s *model*'s score **then**
27.          *O_Bee*'s *model* ← *new_model*
28.       **else**
29.          *O_Bee*'s *model*'s *exhaustion* += 1
30.    **if** min{*models'* validation losses} < *l_thresh* **then**
31.       **break**
32. *final_models* ← *final_models* ∪ *models*

The mutant models generated could present incongruent structures, such as consecutive convolutional layers with no activation function in between. This was avoided by pre-checking every new architecture proposed: activation functions are added in between convolutional layers, same-type contiguous pooling layers are combined, and activation functions before pooling layers swap positions to optimize the number of operations to be performed.

The models are trained until convergence in order to properly compare their performance on the validation partition. Each model generated is then assigned a score based on the Artificial Bee Colony fitness value, given by $(1+L)^{-1}$, where $L$ is the loss value in the validation dataset. This score is used to compare the original and mutated models and discard the worst performing one, incrementing or resetting the exhaustion counters accordingly. Each Onlooker Bee then selects a model from those that remain based on their scores. Similarly, to the Employed Bees, each Onlooker Bee copies and mutates its chosen model, compares it with the original one, and discards the worse of the two. Some models might not be selected, whereas others could be selected by several Onlooker Bees within a single iteration. This drives the exploration-exploitation tradeoff to lean towards the latter for the most promising regions in the hyperspace. At this point the exhaustion counters of each solution are updated again and a new iteration begins with the Employed Bees.

The hyperparameters of the Chimera Algorithm are: the stopping criteria —either a maximum number of iterations or a threshold for the objective loss function—, the probability of performing each mutation type —that is, adding, subtracting or modifying each kind of layer—, the population size, the model exhaustion limit and, optionally, the bounds to the model hyperparameters' space —the maximum number and types of layers, kernel sizes or strides, or the range of learning rates—. The stopping criteria and the search bounds are the ones that most significantly affect the quality of the models produced. The probability of each mutation type, population size and model exhaustion limit define a preferred direction and average search speed throughout the hyperspace. For instance, 1) keeping a higher probability of removing layers rather than adding them yields a search focused on decreasing model complexity, or 2) using a small population size with a high exhaustion limit favors a thorough exploitation of a few regions rather than widespread exploration. The optimal hyperparameters will be problem-dependent, and could be either fixed or adaptive, leveraging a-priori knowledge provided with the quality of the models found throughout the search.

3. EVALUATION

The Chimera Algorithm was evaluated in two scenarios fully described in sections 3.1 and 3.2. In the first scenario, we generated models from scratch to deal with the classification problem of the natural images in the CIFAR-10 [15] dataset, one of the most widely employed datasets for machine learning and computer vision research. In the second scenario, we used the Chimera Algorithm to optimize a given architecture to tackle a problem of biomedical interest: the estimation of the horizontal misalignment of the detector in a Computed Tomography system by analyzing the artifacts present in the reconstructed volumes from a set of projections spanning an angle of 180°.

In both scenarios, four structural hyperparameters were optimized: number of layers, their type —convolutional, max pooling or average pooling—, kernel sizes, and paddings. The model length was only limited by the GPU memory. Convolutional and pooling kernel sizes for each dimension were drawn from a uniform distribution (from 1 to 7). The probabilities of adding, deleting, or mutating a layer on each exploration step were set to 30% each, while there was a 10% probability of simply resetting the weights of some layers while leaving the architecture intact. When adding or mutating a layer, its probability of being a convolutional layer was set as $5n_p/5n_p + n_c$, where n_p is the number of pooling layers and n_c that of convolutional ones in the model to mutate. Otherwise, the new layer is a pooling one, with equal probability of being either max or average pooling. This ensures that our models will tend to present 5 times as many convolutional layers as pooling ones. The learning rate, a hyperparameter commonly optimized in the Neuroevolution field, was instead approximated through the Leslie Smith's learning rate test [16], as it proved itself much faster and reliable than adding an extra dimension to the hyperparameter space. The code developed can be accessed through the project's GitHub repository.

All test were performed with an Intel® Core™ i7-7700 CPU and a NVIDIA® GeForce® RTX 2060 Super™ GPU.

**3.1. Scenario 1: CIFAR-10 dataset**
In this scenario, the Chimera Algorithm was initially run with a population size of 4 employed bees and a search length of 16 iterations to check the capabilities of the algorithm to find good performing models in relatively shallow searches. Afterwards, it was run with a population size of 8 employed bees and a search length of 32 iterations to measure the possible improvement that a lengthier search would yield. The exhaustion limit was set to 10 in both cases. Each test was performed 4 times, reshuffling the training and validation partitions. A learning rate of 10-3 was fixed, as the Leslie Smith test yielded stable convergence for 16 randomly generated models of varying length for learning rates ranging from $10^{-2}$ to $10^{-4}$.

The top-1 accuracy of the architectures generated was compared with that of two —not pretrained— Transfer Learning architectures: Lenet-5 [17], which has around 60000 weights to optimize, and VGG11 [18], with about 133 million parameters. It was necessary to add an extra ReLU and a fully connected layer to the latter to reduce the number of output classes from 1000 in ImageNet to 10 in CIFAR-10. Both experiments were run 32 times, reinitializing the weights and randomly generating a different training and validation partition of the dataset in each iteration. The learning rates employed were 10-3 and 10-4 for the shallow and deep models respectively, as suggested by the Leslie Smith test.

**3.1. Application to the CT misalignments problem**
Horizontal misalignment of the detector in a Computed Tomography system leads to artifacts in the reconstructed images, which appear as upwards or downwards facing arcs, according to the direction of the misalignment (Fig. 1). The greater the misalignment, the thicker the artifact produced.

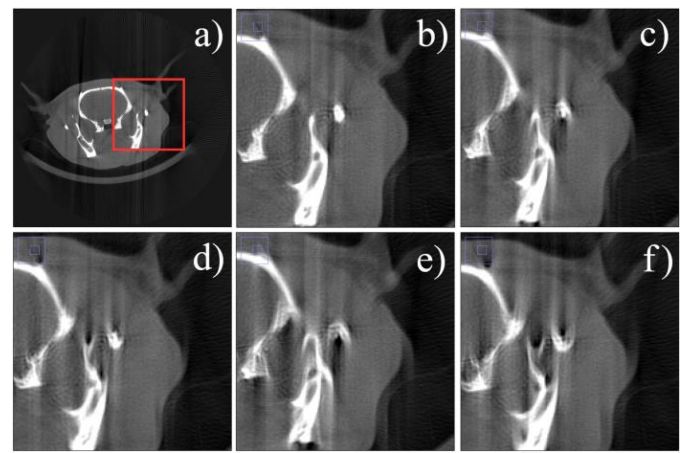

Fig. 1. Axial view of a rodent head CT study reconstructed with an angular span of 180° and with no geometric misalignments (a). Detail of the region indicated in red reconstructed with a simulated horizontal detector misalignment of 0mm (b), +0.5mm (c), -0.5mm (d), +1mm (e) and -1mm (f).

In this experiment we used six rodent cranial studies obtained with a SEDECAL micro-CT system [19] to simulate 25 sets of miscalibrated projections for each volume, by means of the FUX-Sim [20] simulation software. Simulated horizontal misalignments values were taken from a random uniform distribution with a range of ±1mm. An FDK-based algorithm was used for reconstruction, generating volumes of 256×256×200 pixels, which were then normalized to ImageNet's mean and standard deviation —0.485 and 0.225, respectively— and separated into 2D axial slices. The tolerance, defined as the smallest misalignment that produces artifacts noticeable to the naked eye, was set to ±0.1 millimeters [21]. The database was split into a training set with four rodents —20000 images— and validation and test sets with 1 rodent — 5000 images— each.

Four Convolutional Neural Networks were generated with the Chimera Algorithm using the VGG11 [18] architecture as a starting point for the evolution and a search length of seven iterations. The Huber Loss [22] was used as the loss function for the prediction of the detector horizontal misalignment for every slice during training. The $\delta$ value in the Huber Loss, which controls the slackness for outliers, was selected via grid-search within a range [0, 0.5] in 0.1 intervals. The validation performance of VGG11 yielded 0.1 as the best $\delta$ value, which is consistent with the tolerance value defined and was employed for training all models. A suitable learning rate had to be calculated before training every model using the Leslie Smith test, as in this case it did not produce a stable learning rate for differently sized randomly generated models. To reduce the time required to obtain such learning rate, the search was restricted to an order of magnitude above and below that of the parent model to the one to be trained, relying on the assumption that small changes in the architecture would not yield great fluctuations in the optimum learning rate. No exhaustion limit was specified. The Chimera Algorithm's output models are referred to as $VGG_{EV}$.

The misalignment prediction for each volume is given by the median of the predicted misalignments for each slice. The use of the median reduces the influence of outliers coming from slices with less information. The performance of the architectures, measured by the MSE of this median prediction with respect to the true value, was compared with that of four representative members of the VGG transfer learning family proposed in [18] —VGG11, VGG13, VGG16 and VGG19—, as these are some of the highest performing and more widely studied feedforward neural networks employed in image processing. These models had to be slightly modified to fit the expected input and output sizes. An extra layer was added to the fully connected classifier at the rear end of each model to obtain a single regression value instead of the original 1000 class probabilities, and an extra convolutional layer with an output dimension of 3 and a kernel size, stride and weights of 1 was added to the front end to transform the grayscale slices into 3-channeled ones. The learning rate used for each model was the optimum one suggested by the Leslie Smith test.

## 4. RESULTS

This section presents the results obtained for the CIFAR-10 image classification task (4.1) and the CT geometrical misalignment regression task (4.2).

### 4.1. Results on the CIFAR-10 dataset

Table I shows the crossvalidation statistics and the best accuracy attained by the two Transfer Learning architectures tested. VGG11 attains better accuracy in training, validation, and testing throughout, as shown in bold.

TABLE I
LENET-5 AND VGG11 STANDALONE ACCURACIES

| Model | Train acc. | Val acc. | Test acc. | Training time (min) |
|---|---|---|---|---|
| Lenet-5 | 68.96 ± 1.73 (max 72.55) | 60.52 ± 0.76 (max 62.15) | 60.52 ± 0.83 (max 62.37) | 9.988 ± 2.475 |
| VGG11 | **84.01 ± 3.22 (max 88.32)** | **69.15 ± 1.04 (max 71.07)** | **68.82 ± 1.14 (max 70.65)** | 13.751 ± 0.652 |

Table II shows the accuracy of the models generated with the Chimera Algorithm throughout all crossvalidation partitions. The best models were always found in the deep searches. However, some low performing models also appear in deep searches due to the late exhaustion and reinitialization of some solutions, which lowers their mean population accuracy. The time required and the number of models generated in the deep searches were approximately 4-fold that of the shallow searches, which was consistent with the batch size and search length.

TABLE II
CHIMERA ALGORITHM'S OUTPUT MODELS' ACCURACIES

| Tests | Train acc. | Val acc. | Test acc. | Search time (h) | Number of models |
|---|---|---|---|---|---|
| Shallow searches | **85.67 ± 2.84** (max 90.48) | **74.52 ± 0.91** (max 76.62) | **74.43 ± 0.97** (max 76.35) | 8.041 ± 0.226 | 5.25 ± 1.30 |
| Deep searches | 85.07 ± 4.73 (**max 94.66**) | 74.08 ± 4.02 (**max 78.50**) | 73.74 ± 4.00 (**max 78.08**) | 36.23 ± 6.06 | 23.00 ± 4.32 |

### 4.2. Results on the CT misalignments dataset

Deeper Transfer Learning models resulted in better and more reliable results than shallower ones, as shown in Table III.

TABLE III
TRANSFER LEARNING MODELS' MEDIAN PREDICTIONS' ABSOLUTE ERROR ON THE CT ARTIFACTS DATASET (IN MILLIMETERS)

| Models | Train | Val | Test |
|---|---|---|---|
| VGG11 | 0.018 ± 0.018 | 0.316 ± 0.278 | 0.469 ± 0.358 |
| VGG13 | 0.101 ± 0.065 | 0.306 ± 0.223 | 0.264 ± 0.149 |
| VGG16 | 0.009 ± 0.007 | 0.183 ± 0.175 | 0.241 ± 0.150 |
| VGG19 | 0.012 ± 0.009 | 0.124 ± 0.060 | 0.150 ± 0.134 |

Not only do they obtain better median misalignment predictions, but they also present less intra-volume variation for different slices, as can be appreciated in Fig. 2

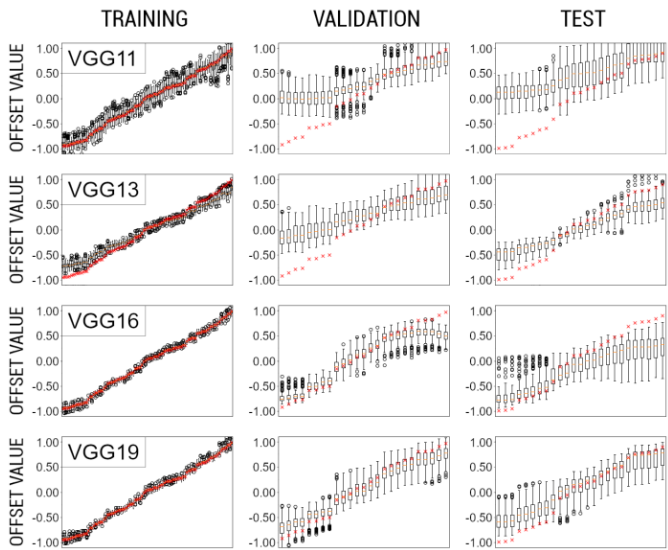

Fig. 2. Slicewise misalignment values predicted by VGG11, VGG13, VGG16 and VGG19 (from top to bottom, respectively) for each volume. The boxplots are ordered along the x axis based on their target misalignment value (in red).

Fig. 3 shows the evolution of the training, validation, and test performances of the best model throughout the Chimera Algorithm's search. Even though the median predictions did not always change considerably from one iteration to the next — especially in training, as observed in Fig. 3—, the resulting models became much more reliable, presenting a much slower standard deviation for the misalignment prediction for slices within the same volume in Fig. 4.

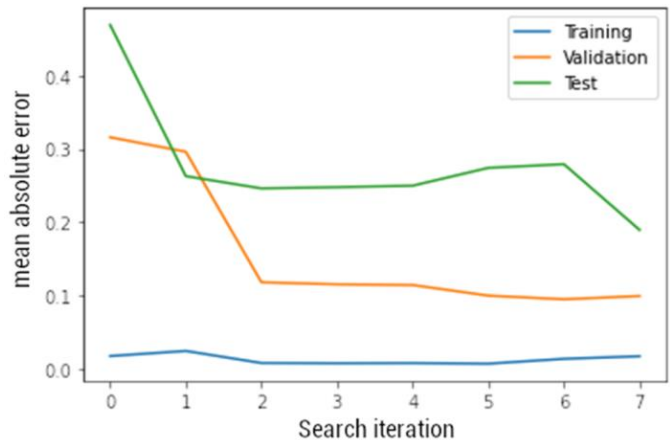

Fig. 3. Evolution of the performance of the best model found on each iteration of the Chimera Algorithm.

The performance of all 4 final models generated is shown in Table IV. The complete outputs of these models can be seen in Fig. 5. All VGGEV models performed better than the original VGG11 in all partitions. The best model generated, VGGEV 1, reduced the error attained with VGG11 by 60% in the test dataset. Although some of the VGGEV models surpassed the performance of the deeper VGG19 in validation, none of them did so in the testing partition.

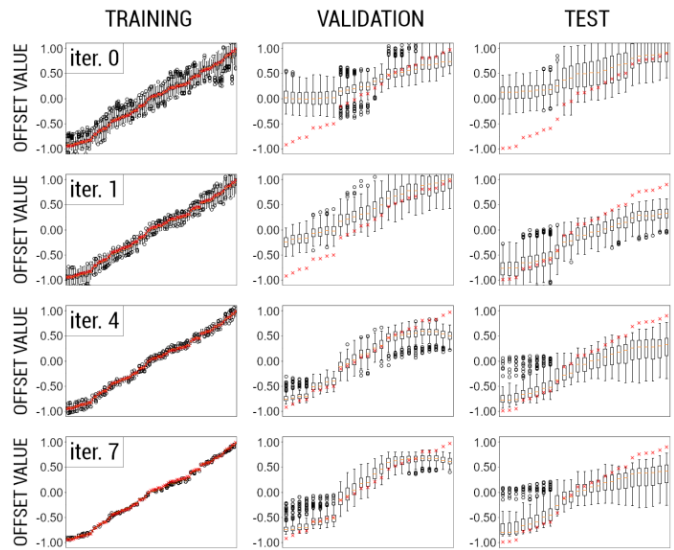

Fig. 4. Slicewise misalignment values predicted by VGG11 (first row) and the best VGGEV found at iterations 1, 4, and 7 (second to fourth rows) for each volume. The boxplots are ordered along the x axis based on their target misalignment value (in red).

TABLE IV
CHIMERA ALGORITHM GENERATED MODELS' MEDIAN PREDICTIONS'
ABSOLUTE ERROR ON THE CT ARTIFACTS DATASET (IN MILLIMETERS)

| Models | Train | Val | Test |
|---|---|---|---|
| $VGG_{EV}$ 0 | 0.013 ± 0.010 | 0.107 ± 0.102 | 0.254 ± 0.151 |
| $VGG_{EV}$ 1 | 0.017 ± 0.013 | 0.099 ± 0.083 | 0.189 ± 0.127 |
| $VGG_{EV}$ 2 | 0.013 ± 0.006 | 0.104 ± 0.101 | 0.224 ± 0.114 |
| $VGG_{EV}$ 3 | 0.051 ± 0.038 | 0.130 ± 0.134 | 0.254 ± 0.136 |

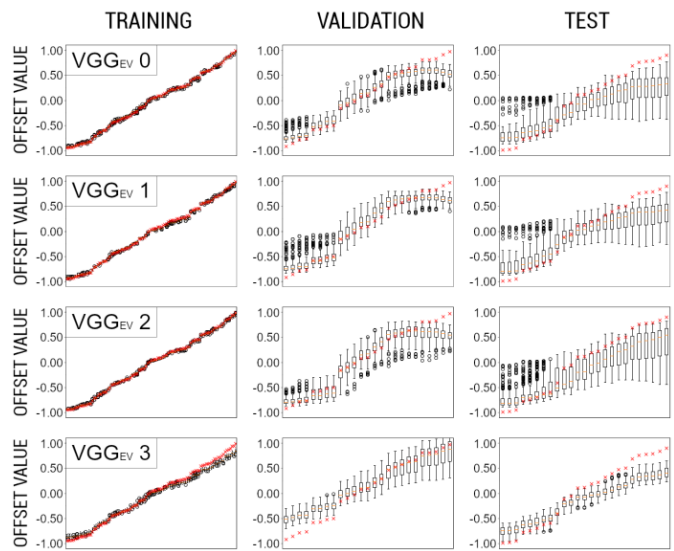

Fig. 5. Slicewise misalignment values predicted by the models generated by the Chimera Algorithm for each volume. The boxplots are ordered along the x axis based on their target misalignment value (in red).

5. DISCUSSION AND CONCLUSIONS

This work presents the Chimera Algorithm, a novel optimizer able to generate and evolve the architecture of feedforward

Convolutional Neural Networks based on the adaptation of the Artificial Bee Colony Algorithm to work within an Evolutionary Computation framework.

The evaluation showed that the Chimera Algorithm surpassed the performance of Transfer Learning in the training and validation partitions in both scenarios considered. However, VGG19 was able to outperform the generated models in the test partition of the CT artifacts dataset. This was the only case where a pretrained model was able to generalize better than the fine-tuned architectures, even though the latter outperformed VGG19 in both training and validation partitions. This suggests that the Chimera Algorithm might be overfitting on these partitions, thus hindering the generalization capabilities of the resulting models. This happens because the validation dataset is repeatedly employed throughout the search to select or discard the models. The problem may become paticularly severe with small dataset sizes —such as in our case—, for which is difficult to create representative training and validation datasets. The model's ability to generalize could be enhanced by employing data augmentation [23] o validating the architectures using aditional test dataset partitions throughout the search. Moreover, considering the uncertainty of the models predictions, as quantified through methods such as those explored in the literature [24], would allow for a well-founded pruning the population of solutions, mitigating the impact of overfitting. Another possible option to improve model generalization could be to combine the populations of generated architectures into ensembles, which has already been shown to reduce bias and variance of individual learners [25].

The Chimera Algorithm is able to overcome the premature convergence problem, typical of traditional Genetic and Evolutionary-based Neuroevolution [9], as non-promising solutions are not discarded throughout the search, but rather just given less attention. This allows the preservation of great diversity within the population, which would in turn be highly beneficial for its combination into ensembles, as achieving high model diversity is one of the main challenges of Deep Ensemble Learning [26]. Moreover, as discussed previously, other Swarm Intelligence approaches like Particle Swarm Optimization lack the ability to exploit interesting regions locally nor reach faraway solutions. This makes them very dependent on the initialization. Some techniques developed to overcome these limitations, such as the introduction of momentum or constraint factors [27] are not defined for combinatorial problems. The Chimera Algorithm does not require the use of these techniques as it exploits the local neighboring solutions via mutation operators and is free to explore further away as the solutions become exhausted, allowing for both the creation of novel models from scratch and the optimization of given architectures provided by the user. Moreover, unlike in supernet optimization approaches [10] or Multi-Fidelity MetaLearning [11], the solution space is only bounded by hardware limitations. Despite the wide size of the solution space. The algorithm is able to converge to adequate solutions in a reasonable period of time. Other sophisticated approaches, such as training a Controller Recurrent Neural Network using Reinforcement Learning to generate the architectures [28] achieved considerably better results in CIFAR-10 —up to 96.35 accuracy in the test partition—, but they do so by sheer brute force, requiring 28 days and 800 K40 GPUs, compared to the few dozen hours of our approach on a RTX 2060 Super GPU.

In conclusion, in this work we show the potential of the Artificial Bee Colony Algorithm as a backbone for the optimization of feedforward Convolutional Neural Network architectures. The Chimera Algorithm can be applied straight away to regression or clasification problems, and generates suitable models in an unsupervised way and reasonable amounts of time. As such, it may pave the way to create even more powerful optimization techniques in an automatic and unsupervised way.

REFERENCES

1. A. Esteva, A. Robicquet, B. Ramsundar et al., "A guide to deep learning in healthcare," *Nat Med*, vol. 25, pp. 24-29, Jan. 2019. DOI: 10.1038/s41591-018-0316-z.
2. M. Raghu, C. Zhang, J. Kleinberg and S. Bengio, "Transfusion: Understanding Transfer Learning for Medical Imaging," in *NEURIPS 2019*, Vancouver, Canada, 2019.
3. A. Klein, S. Falkner, J. T. Springenberg and F. Hutter, "Learning Curve Prediction with Bayesian Neural Networks," in *ICLR 2017*, Toulon, France, Apr. 24-26, 2017.
4. C. Jiang et al., "Neural Capacitance: A New Perspective of Neural Network Selection via Edge Dynamics," *arXiv*, Jan. 2022. DOI: https://doi.org/10.48550/arXiv.2201.04194, [Online].
5. A. Baldominos, Y. Sáez and P. Isasi, "On the Automated, Evolutionary Design of Neural Networks-Past, Present, and Future," *Neural Computing and Applications*, vol. 32, pp. 519-545, 2020. DOI: 10.1007/s00521-019-04160-6.
6. S. Harp, T. Samad and A. Guha, "Designing application specific neural networks using the genetic algorithm," *Advances in neural information processing systems*, vol. 2, pp 447–454, 1989.
7. H. Kitano, "Designing neural networks using genetic algorithms with graph generation system," *Complex Syst*, vol. 4, pp. 461–476, 1990.
8. E. Ronald and M. Schoenauer, "Genetic Lander: An experiment in accurate neuro-genetic control", *PPSN III 1994 Parallel Programming Solving from Nature*, pp. 452–461, 1994. DOI: 10.1.1.56.3139
9. H. Iba, *Swarm Intelligence and Deep Evolution: Evolutionary Approach to Artificial Intelligence*, 1$^{st}$ ed., pp. 63-65, 104-132, Apr. 2022, ISBN 9781032009155.
10. S. Cha et al., "A Survey of Supernet Optimization and its Applications: Spatial and Temporal Optimization for Neural Architecture Search," arXiv preprint arXiv:2204.03916, 2022.
11. L. Zimmer, M. Lindauer and F. Hutter, "Auto-Pytorch: Multi-Fidelity MetaLearning for Efficient and Robust AutoDL," in *IEEE Transactions on Pattern Analysis and Machine Intelligence*, vol. 43, no. 9, pp. 3079-3090, 1 Sept. 2021, DOI: 10.1109/TPAMI.2021.3067763.
12. B. Wang, Y. Sun, B. Xue, M. Zhang, "Evolving deep convolutional neural networks by variable-length particle swarm optimization for image classification," in *2018 IEEE Congress on Evolutionary Computation (CEC)*, pp. 1514–1521, 2018.
13. J. Fernandes, E. Francisco and G. Yen, "Particle swarm optimization of deep neural networks architectures for image classification," *Swarm and Evolutionary Computation*, vol. 49, pp. 62-74, 2019. DOI: 10.1016/j.swevo.2019.05.010.
14. D. Karaboga and B. Basturk, "A powerful and Efficient Algorithm for Numerical Function Optimization: Artificial Bee

Colony (ABC) Algorithm," *Journal of Global Optimization*, vol. 39, pp 459-471, Apr. 2017, DOI: 10.1007/s10898-007-9149-x.
15. A. Krizhevsky and G. Hinton, "Learning multiple layers of features from tiny images," *M.S. Thesis, Dept. of Computer Science, University of Toronto,* Toronto, 2009.
16. L. N. Smith, "A disciplined approach to neural network hyper-parameters: Part 1 -- learning rate, batch size, momentum, and weight decay," *arXiv Computing Research Repository*, Mar. 2018.
17. Y. Lecun, L. Bottou, Y. Bengio and P. Haffner, "Gradient-based learning applied to document recognition," in *Proceedings of the IEEE*, vol. 86, no. 11, pp. 2278-2324, Nov. 1998, DOI: 10.1109/5.726791.
18. K. Simonyan and A. Zisserman, "Very Deep Convolutional Networks for Large-Scale Image Recognition," in *ICLR 2015*, San Diego, CA, USA, May 7-9, 2015.
19. J. J. Vaquero et al., "Assessment of a New High-Performance Small-Animal X-Ray Tomograph," *IEEE Transactions on Nuclear Science*, vol. 55, no. 3, pp 898-905, Jun. 2008, DOI: 10.1088/0031-9155/54/18/005.
20. M. Abella et al., "FUX-Sim: Implementation of a fast universal simulation/reconstruction framework for X-ray systems," *PLOS ONE*, vol. 12, no. 7, pp. 1–22, Jul. 2017, DOI: 10.1371/journal.pone.0180363.
21. M. Abella et al., "Tolerance to geometrical inaccuracies in CBCT systems: A comprehensive study," *Med. Phys.*, 2021. DOI: 10.1002/mp.15065
22. P. J. Huber, "Robust Estimation of a Location Parameter," in *Kotz, S., Johnson, N.L. (eds) Breakthroughs in Statistics, Springer Series in Statistics*, vol. 35, New York, NY, USA, Springer, pp. 492–518, 1992.
23. T. Kumar, A. Mileo, R. Brennan and M. Bendechache, "Image Data Augmentation Approaches: A Comprehensive Survey and Future directions", *arXiv preprint*, 2023. arXiv:2301.02830, [Online].
24. 24. M. Abdar et al., "A review of uncertainty quantification in deep learning: Techniques, applications and challenges," *Information Fusion*, vol. 76, pp. 243-297, 2021. DOI: 10.1016/j.inffus.2021.05.008.
25. L. Breiman, "Bias, variance, and arcing classifiers", *Technical Report, Statistics Dept, University of California*, Berkeley, CA, Apr. 1996.
26. M. A. Ganaie, Minghui Hu, A.K. Malik, M. Tanveer and P.N. Suganthan, "Ensemble deep learning: A review," *Engineering Applications of Artificial Intelligence*, vol. 115, Aug. 2022, DOI: https://doi.org/10.48550/arXiv.2104.02395, [Online].
27. M. Imran, R. Hashim and N. E. A. Khalid, "An Overview of Particle Swarm Optimization Variants," *Procedia Engineering*, vol. 53, pp. 491-496, 2013. DOI: 10.1016/j.proeng.2013.02.063.
28. 28. Zoph and Q. V. Le, "Neural Architecture Search with Reinforcement Learning," in *ICLR 2017*, Toulon, France, 2017.